\documentclass[pra,twocolumn,showpacs,amsmath,amssymb,floatfix,superscriptaddress]{revtex4-1}
\usepackage{graphicx}
\usepackage{dcolumn}
\usepackage{bm}
\usepackage{color}
\usepackage{graphicx}
\usepackage{epstopdf}
\usepackage{epsfig}
\usepackage[colorlinks=true, citecolor=blue, anchorcolor=green]{hyperref}

\include{graphic}

\makeatletter
\begin{document}
\preprint{APS/123-QED}

\title{Control of the magnon-polariton hybridization with a microwave pump}

\author{C.~Zhang}
\affiliation{School of Physical Science and Technology, ShanghaiTech University, Shanghai 201210, China}
\affiliation{Shanghai Institute of Microsystem and Information Technology, Chinese Academy of Sciences, Shanghai 200050, China}

\author{Jinwei~Rao}\email{raojw@shanghaitech.edu.cn}
\affiliation{School of Physical Science and Technology, ShanghaiTech University, Shanghai 201210, China}

\author{C. Y.~Wang}
\affiliation{School of Physical Science and Technology, ShanghaiTech University, Shanghai 201210, China}

\author{Z. J.~Chen}
\affiliation{School of Physical Science and Technology, ShanghaiTech University, Shanghai 201210, China}

\author{K. X.~Zhao}
\affiliation{School of Physical Science and Technology, ShanghaiTech University, Shanghai 201210, China}

\author{Bimu~Yao}\email{yaobimu@mail.sitp.ac.cn;}
\affiliation{School of Physical Science and Technology, ShanghaiTech University, Shanghai 201210, China}
\affiliation{State Key Laboratory of Infrared Physics, Shanghai Institute of Technical Physics, Chinese Academy of Sciences, Shanghai 200083, China}

\author{Xu-Guang~Xu}
\affiliation{School of Physical Science and Technology, ShanghaiTech University, Shanghai 201210, China}

\author{Wei~Lu}\email{luwei@mail.sitp.ac.cn;}
\affiliation{School of Physical Science and Technology, ShanghaiTech University, Shanghai 201210, China}
\affiliation{State Key Laboratory of Infrared Physics, Shanghai Institute of Technical Physics, Chinese Academy of Sciences, Shanghai 200083, China}

\begin{abstract}
Pump-induced magnon modes (PIMs) are recently discovered elementary excitations in ferrimagnets that offer significant tunability to spin dynamics. Here, we investigate the coupling between a PIM and cavity magnon polaritons (CMPs) by driving a cavity magnonic system away from equilibrium with a microwave pump. In our experiment, the Walker mode simultaneously couples with the PIM and cavity photons and thus combines two strongly coherent coupling processes in a single cavity structure. Such a PIM-CMP hybridization system acquires complementary properties from both the PIM and CMPs, allowing it to be freely manipulated by the magnetic field, the pump power and the pump frequency. These coherent manipulations exhibit unique behaviors beyond the intrinsic properties limited by the material nature and electromagnetic boundary conditions, thereby creating opportunities for extending the control of hybrid devices. 

\end{abstract}

\maketitle

\section{Introduction}
The last decade has seen the rapid development of the fruitful field of cavity magnonics \cite{soykal2010strong,huebl2013high,tabuchi2015coherent,goryachev2014high,lachance2019hybrid,bhoi2020roadmap,cao2015exchange,boventer2018complex,yu2019prediction,grigoryan2018synchronized,rao2021interferometric,li2022coherent,chelpanova2021intertwining}. Because of its strong potential in quantum information processing and spintronics, this field of research has attracted increasing attention. A cavity magnonic system is typically composed of microwave cavities and bulk magnets \cite{huebl2013high,tabuchi2015coherent}.  Magnons, i.e., the engergy quanta of spin waves, in magnets can strongly interact with microwave photons in cavities, producing quasi-particles named ``cavity magnon polaritons'' (CMPs). These quasi-particles acquire complementary properties from cavity photons and magnons, and thus are highly tunable and have long coherence\cite{zhang2015magnon,Tobar2014,yu2020}, diverse non-linearities\cite{BSCMP,TSCMP,Paul2018}, and excellent compatibility\cite{NK2016,xu2021,AFM2021}. By exploiting these merits, numerous techniques and applications have been developed, such as long time memories \cite{zhang2015magnon,shen2021long}, polaritonic logic gates \cite{rao2019analogue}, magnon-phonon entanglement \cite{li2018magnon}, spin current manipulation \cite{bai2017cavity}, and information processing at the quantum limit \cite{lachance2020entanglement,tabuchi2015coherent,tabuchi2014hybridizing}. All these applications rely on an effective access to manipulate the photon-magnon coupling, which is usually implemented by delicately designing the microwave cavities.

\begin{figure} [h!]
\begin{center}
\epsfig{file=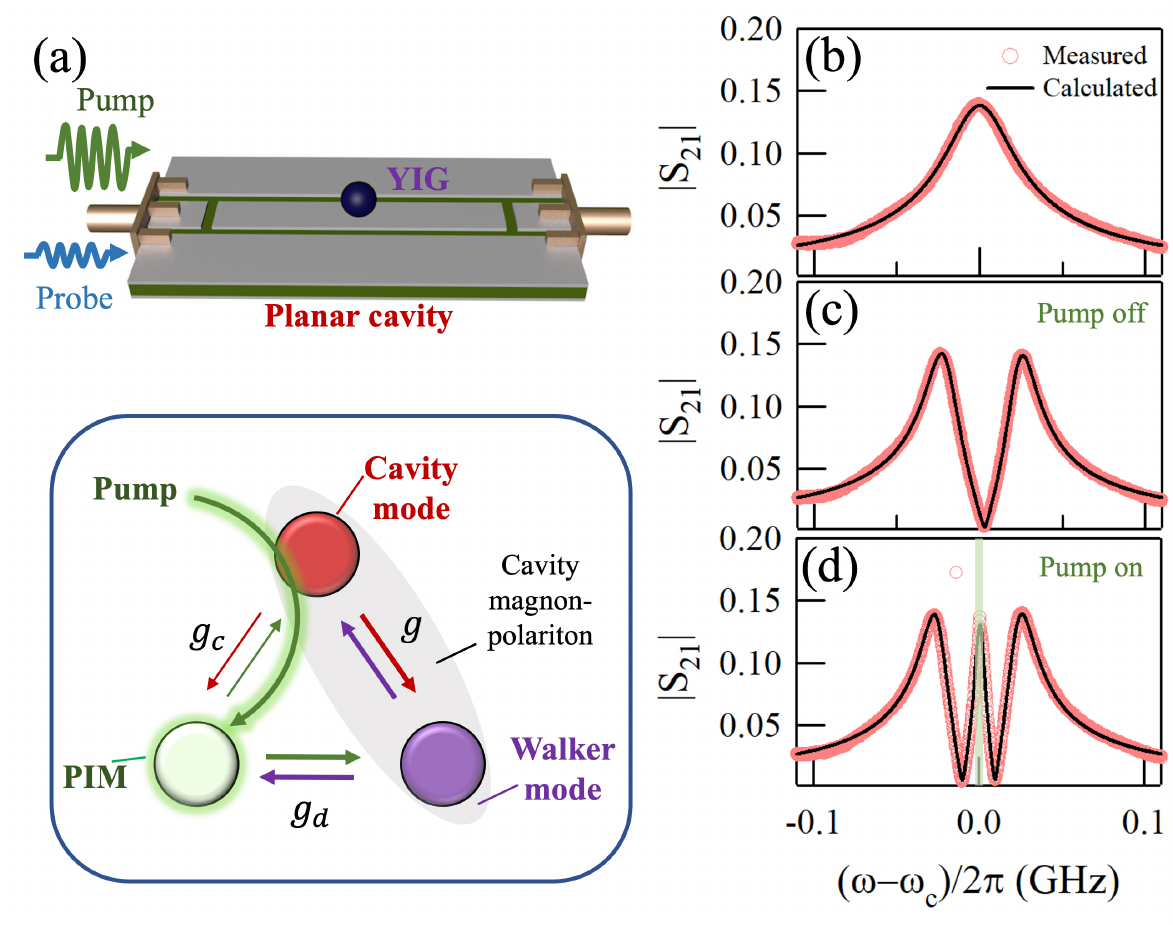,width=8.5cm} \caption{(a) Upper panel shows a schematic diagram of a cavity magnonic device, which simply consists of a CPW cavity and a YIG sphere. Two tones (i.e., pump and probe) drive and detect the resonances in this device. Lower panel shows a schematic drawing of the PIM-CMP hybridization. (b) Transmission spectra of empty cavity. (c), (d) Transmission spectra of the device without and with a pump. During these two measurements, the Walker mode matches the cavity mode( i.e., $\omega_w=\omega_c$). The pump is set as $\omega_d/2\pi=\omega_c/2\pi=3.34$ GHz and $P_d=10$ dBm. The green strip in panel(d) indicates the pump. The black curves in panels(b)-(d) are the  results of calculations using Eq. (\ref{Trans}).}\label{fig1}
\end{center}
\end{figure}

A recent work has experimentally demonstrated that a ferrimagnet driven by a strong microwave can exhibit non-trivial properties, in particular the emergence of the pump-induced magnon mode (PIM) \cite{RaoPhysRevLett}. The formation of this peculiar magnon mode is attributed to the cooperative precession of unsaturated spins in the ferrimagnet under a microwave pump. The inherent nature of PIM allows it to be highly tunable by the microwave pump. Such a property is not possessed by any other magnon modes in magnets, including magnetostatic modes or Damon-Eshbach modes. If we incorporate a PIM into cavity magnonics, one straightforward consequence is that the cavity magnonic system will acquire an additional manipulation freedom inherited from the PIM and hence become tunable by the microwave pump. A similar technique that uses a strong microwave to manipulate a on-chip cavity magnonic device in the nonlinear regime has been demonstrated \cite{PhysRevLett.123.107701}. Comparing to conventional cavity magnonic devices that highly depend on the design of microwave cavities, a tuning technique that merely needs to adjust the power or frequency of a microwave pump and can work in both the linear and nonlinear regimes is very useful. Moreover, the PIM can strongly couple to normal magnon modes in magnets via spin-spin interactions. This effect makes possible the hybridization of the PIM and CMP, and may hence produce new quasi-particles in cavity magnonics. Combined with the flexibility of PIM dynamics, such a quasiparticle would provide an ideal solution and significant impact on the high-dimensional tunability of the cavity magnonic system.

In this work, we introduce a PIM into cavity magnonics and investigate the strong interaction between the PIM and CMPs. The magnon mode commonly used to produce CMPs is the Walker modes \cite{walker1957magnetostatic,dillon1957ferrimagnetic,gloppe2019resonant}, i.e., a kind of magnetostatic modes in bulk magnets. Its effective spin number is constant determined by the saturated magnetization and volume of the bulk magnet, while its mode frequency is governed by the well-known Kittel formula that is a function of the effective magnetic field felt by a Walker mode. By contrast, the PIM differs in that its effective spin number and mode frequency are tunable via the pump. As a consequence, the PIM-CMP coupling strength $g_d$ follows a fourth root of the pump power (i.e., $g_d\propto P_d^{1/4}$). It can also be controlled via the detuning between the pump and the cavity mode because this detuning regulates the effective field intensity on the intra-cavity yttrium iron garnet (YIG) sphere. These results demonstrate the control of a cavity magnonic system with a microwave pump. This technique enriches the tunability of cavity magnonic systems and is useful for the magnon-based information processing. Attaining a better understanding of the nontrivial properties of the PIM-CMP interaction will be beneficial for the development of coherent/quantum information processing based on cavity magnonics. 

\begin{figure} [h!]
\begin{center}
\epsfig{file=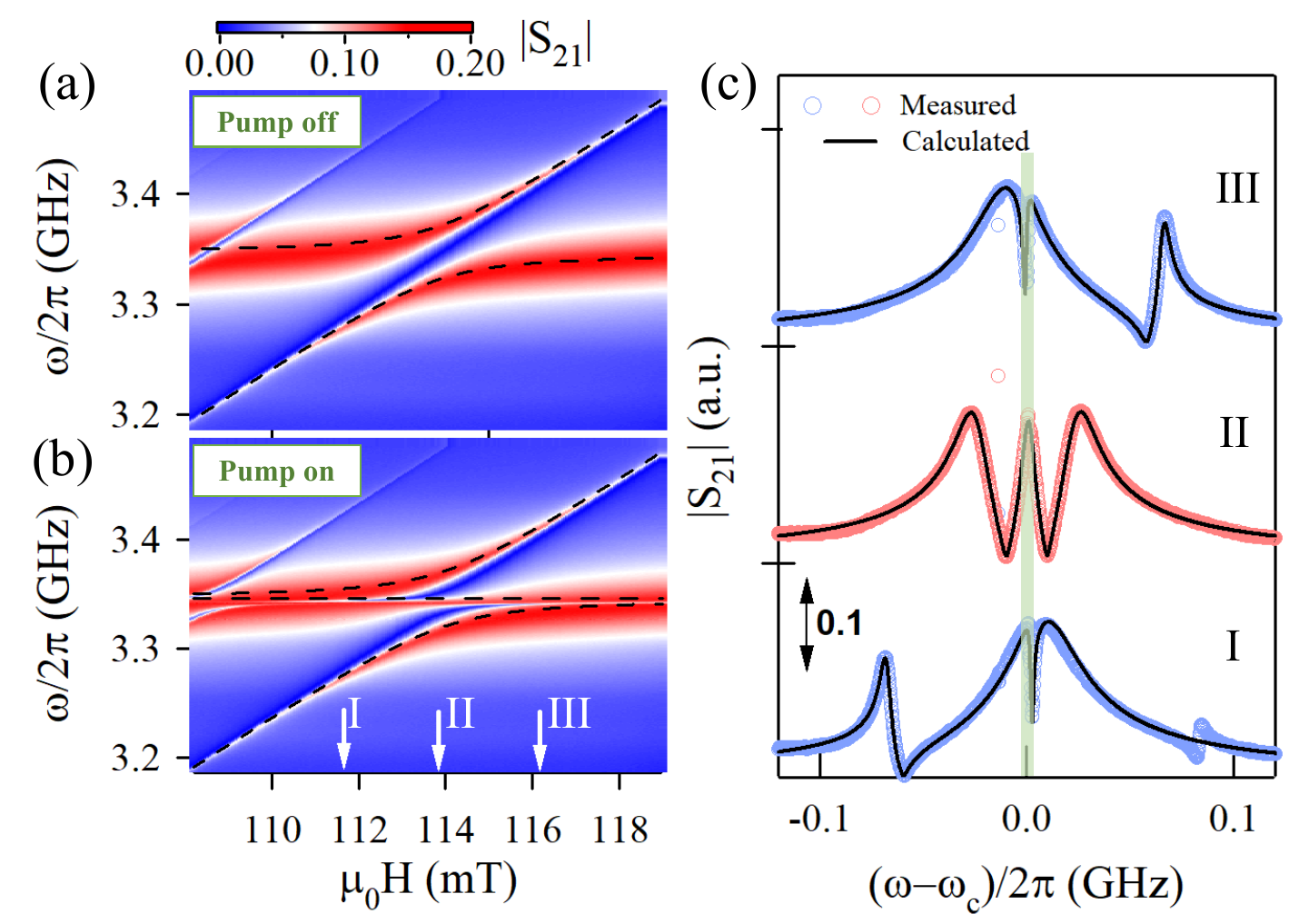,width=8.5cm} \caption{Transmission spectra of our device measured at different magnetic fields (a) without  and (b) with  the pump, which is set at $\omega_d=\omega_c$ and $P_d=10$ dBm. Black dashed lines are eigenfrequencies calculated from the Hamiltonian in two cases. The anticrossing at 100 mT arises from the coupling between another Walker mode and the cavity mode. (c) Three transmission spectra measured at $\mu_0H_{ext}=$111.7, 113.9 and 116.1 mT [indicated by white arrows in panel (b)] with the pump on. Black solid lines are transmission spectra calculated by using Eq. (\ref{Trans}).}\label{fig2}
\end{center}
\end{figure}

\section{Experiment}
Figure\,\ref{fig1}(a) shows the schematic picture of a simplest cavity magnonic device, which consists of a coplanar waveguide (CPW) cavity and a polished YIG sphere of diameter 1 mm. The CPW cavity is fabricated on a $22\times50$~mm$^2$ RO4350b board that contains a hydrocarbon ceramic laminate (0.76 mm) and two copper layers (0.035 mm). The fundamental mode of the CPW cavity is at $\omega_c/2\pi=3.34$ GHz. Its external and intrinsic damping rates fitted from the transmission spectrum [Fig.\,\ref{fig1}(b)] are $\kappa/2\pi=2.9$ MHz and $\beta/2\pi=16.8$ MHz, respectively. A strong microwave pump generated by a microwave generator is used to drive the system. It is a single-frequency signal at the frequency $\omega_d$. A weak probe is emitted and collected by a vector network analyser, whose power is fixed at $-25$ dBm throughout this work and whose frequency ($\omega_p$) is swept over a wide range for the transmission measurement. All measurements are performed at room temperature.

An external magnetic field $H_{ext}$ parallel to the plane of the CPW is used to tune the Walker modes ($\omega_w$) in the YIG sphere. When a Walker mode, i.e. (2,2,0) mode in this work, matches the cavity mode (i.e., $\omega_w=\omega_c$), the strong photon-magnon coupling produces CMPs. The transmission spectrum [Fig.\,\ref{fig1}(c)] shows two resonance peaks with a Rabi gap, corresponding to two CMP eigenstates. We then turn on the pump and set its frequency to $\omega_d=\omega_c$ and its power to $P_d=10$ dBm. The strong pump drives the unsaturated spins in the YIG sphere via the cavity. The collective precession of these unsaturated spins forms a spin wave at the frequency $\omega_d$, namely the PIM \cite{RaoPhysRevLett}. The PIM has much fewer spins than Walker modes, so it couples negligibly with the cavity mode. However, the PIM can strongly couple to the Walker mode via the spin-spin interaction, which is orders of magnitude stronger than the magnetic dipole-dipole interaction between a magnet and the rf-field. The PIM-CMP hybridization process is depicted in Fig.\,\ref{fig1}(a) as a three-mode coupling. The transmission spectrum in Fig.\,\ref{fig1}(d) shows three resonant peaks instead of the former Rabi splitting, representing the three eigenstates of the PIM-CMP hybridization.

\begin{figure} [h]
\begin{center}
\epsfig{file=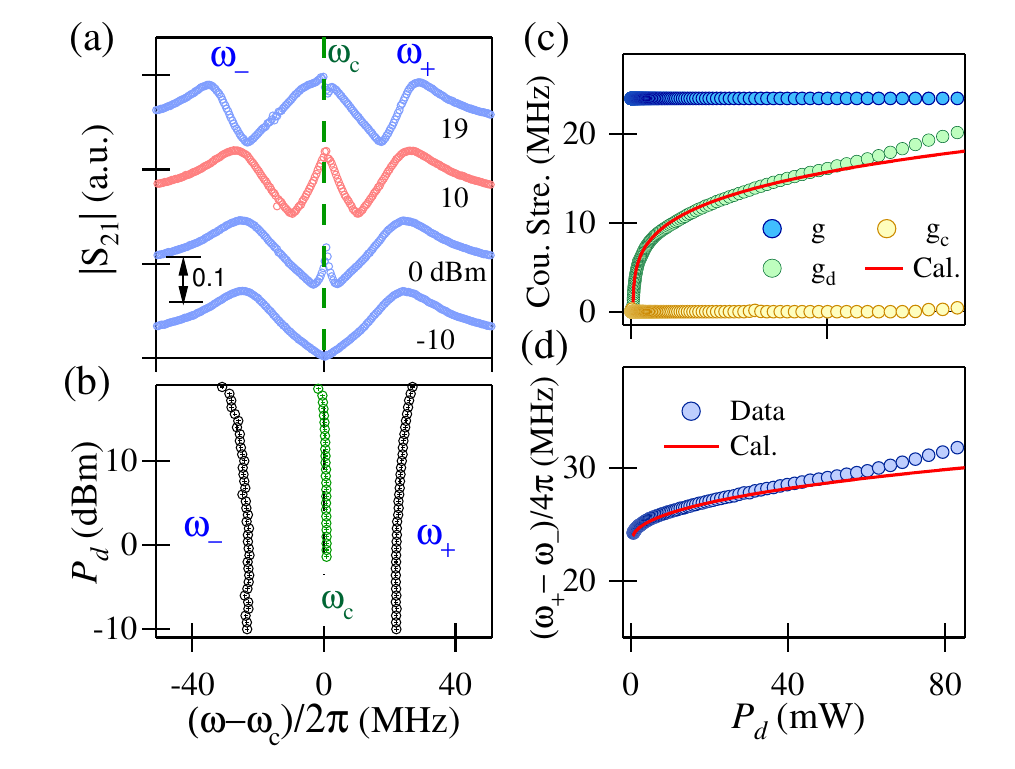,width=8.5cm} \caption{(a) Transmission spectra measured at $P_d=$ -10, 0, 10, and 19 dBm with setting $\omega_w=\omega_d=\omega_c$. (b) Frequencies of three hybridized modes extracted from transmission spectra measured at different pump powers. (c) Three coupling strengths $g$, $g_d$ and $g_c$ fitted from the transmission spectra shown in (a). Red solid line is a calculating curve by using Eq. (\ref{CS0}). (d) Half coupling gaps [i.e., $(\omega_+-\omega_-)/2$] measured at different pump powers. Red solid line is a calculating curve of $\sqrt{g_d^2+g^2}/2\pi$ by using Eq. (\ref{CS0}). }\label{fig3}
\end{center}
\end{figure}

The PIM-CMP hybridization arises from the strong coupling of three resonant modes. To quantitatively model this process, we construct a Hamiltonian model containing three modes and their coupling effects. We focus our research on the coupling effects in our system in the linear regime and thus disregard the self- and cross-Kerr effects \cite{wu2021observation,anderson1955instability}. The Hamiltonian is
\begin{eqnarray}
\mathcal{H}/\hbar&=&\tilde{\omega}_w\hat{a}^\dag\hat{a}+\tilde{\omega}_d\hat{b}^\dag\hat{b}+\tilde{\omega}_c\hat{c}^\dag\hat{c}+g_d(\hat{a}^\dag\hat{b}+\hat{a}\hat{b}^\dag)+\nonumber \\
&\quad&g(\hat{a}^\dag\hat{c} +\hat{a}\hat{c}^\dag)+g_c(\hat{b}^\dag\hat{c}+\hat{b}\hat{c}^\dag)+\nonumber \\
&\quad&[i\sqrt{\kappa}(A_pe^{-i\omega_pt}+A_de^{-i\omega_dt})\hat{c}^\dag +\rm H.C.],
\label{Hamiltonian}
\end{eqnarray}
where $\hat{a}^\dag$ ($\hat{a}$), $\hat{b}^\dag$ ($\hat{b}$) and $\hat{c}^\dag$ ($\hat{c}$) represent the creation (annihilation) operators of the Walker mode, PIM and cavity photon mode, respectively. $\tilde{\omega}_w=\omega_w-i\alpha$, $\tilde{\omega}_d=\omega_d-i\xi$ and $\tilde{\omega}_c=\omega_c-i(\beta+\kappa)$ correspond to their complex frequencies, where $\alpha/2\pi=0.8$ MHz and $\xi/2\pi=0.4$ MHz are the damping rates of the Walker mode and of the PIM, respectively. $A_d$ and $A_p$ are respectively the pump and probe intensities,  which can be simply viewed as the voltage amplitudes of two microwave signals. They follow a square-root relation of microwave power, such as $A_d\propto\sqrt{P_d}$. $g/2\pi=24$ MHz is the coupling strength between the cavity and the Walker mode, larger than the damping rates of all subsystems. $g_d$ is the coupling strength between the Walker mode and the PIM and has the form $g_d=g_0\sqrt{N_p}$, where $g_0$ is a constant and $N_p$ is the effective spin number of the PIM \cite{RaoPhysRevLett,tabuchi2014hybridizing}. Because PIM is induced by the pump, its spin number should be equal to the magnon number excited by the pump, i.e., $N_p=\langle\hat{b}^\dag\hat{b}\rangle$.  $g_c$ is the coupling strength between the PIM and the cavity mode.

Under the two-tone driving, our system oscillates in a superposition of a steady oscillation driven by the strong pump and fluctuations perturbed by the weak probe. we assume that $\hat{a}=[A+\delta u(t)]e^{-i\omega_dt}$, $\hat{b}=[B+\delta v(t)]e^{-i\omega_dt}$ and $\hat{c}=[C+\delta\rho(t)]e^{-i\omega_dt}$, where $A$, $B$ and $C$ respectively represent the steady oscillating amplitudes of three modes at the frequency of $\omega_d$. $\delta u(t)$, $\delta v(t)$ and $\delta\rho(t)$ correspond to their fluctuations induced by the probe. These fluctuations can be derived from Eq. (\ref{Hamiltonian}) (see Appendix A). By using the input-output relation of $S_{21}=\sqrt{\kappa}|\delta\rho(t)|/A_p$, we get the transmission spectrum of our device:
\begin{equation}
S_{21}=-\frac{\kappa}{i(\omega_p-\tilde{\omega}_c)+i\frac{(\omega_p-\tilde{\omega}_d)g^2+(\omega_p-\tilde{\omega}_w)g_c^2+2gg_cg_d}{g_d^2-(\omega_p-\tilde{\omega}_d)(\omega_p-\tilde{\omega}_w)}}.
\label{Trans}
\end{equation}
 It can well reproduce the measured transmission spectra. This expression has three maxima, corresponding to the system's eigenfrequencies. Turning off the pump, the PIM disappears, so that our system degrades to a normal cavity magnonic system with two CMP eigenfrequencies. 

The Walker-PIM coupling strength $g_d$ can also be derived from Eq. (\ref{Hamiltonian}) (see Appendix B). It is determined by a cubic equation with a form of \begin{equation}
{\tilde{\Delta}_c}g_0^2|B|^2B-i\sqrt{\kappa}A_dgg_0|B|-i\xi(g^2-\tilde{\Delta}_c\tilde{\Delta}_w)B=0
\label{CS},
\end{equation}
where $\tilde{\Delta}_w =\tilde{\omega}_w -\omega_d$ and $\tilde{\Delta}_c =\tilde{\omega}_c -\omega_d$ are the complex detunings of the Walker mode and the cavity mode with respect to the pump. If we set $\omega_c=\omega_w=\omega_d$, the Walker-PIM coupling strength becomes
\begin{equation}
    g_d=\sqrt{\frac{\sqrt{\kappa}gg_0A_d-\xi g^2}{\beta+\kappa}-\alpha\xi}
    \label{CS0}.
\end{equation}

We now test the theoretical model from three aspects: the dependence of the PIM-CMP hybridization on the magnetic field $H_{ext}$, the pump power $P_d$ and the pump frequency $\omega_d$. Figure\,\ref{fig2}(a) shows the transmission spectra of our device measured at different $H_{ext}$ without the pump. When the Walker mode is tuned to get close to the cavity mode, a typical anticrossing occurs near 114 mT, indicating the strong photon-magnon coupling and hence the formation of CMPs. We then turn on the pump and set it at $\omega_d=\omega_c$ and $P_d=10$ dBm. A PIM at $\omega_d$ is excited by the pump, producing three hybridized modes from the transmission map [Fig.\,\ref{fig2}(b)]. The black dashed lines in Figs.\,\ref{fig2}(a) and \ref{fig2}(b) are eigenfrequencies of our system in the two cases. Transmission spectra measured at $\mu_0H_{ext}=$111.7, 113.9 and 116.1 mT with the pump are plotted in Fig.\,\ref{fig2}(c), which can be well reproduced by using Eq. (\ref{Trans}). $g_d/2\pi$ fitted to these curves is $9.8$ MHz. 

\begin{figure} [ht]
\begin{center}
\epsfig{file=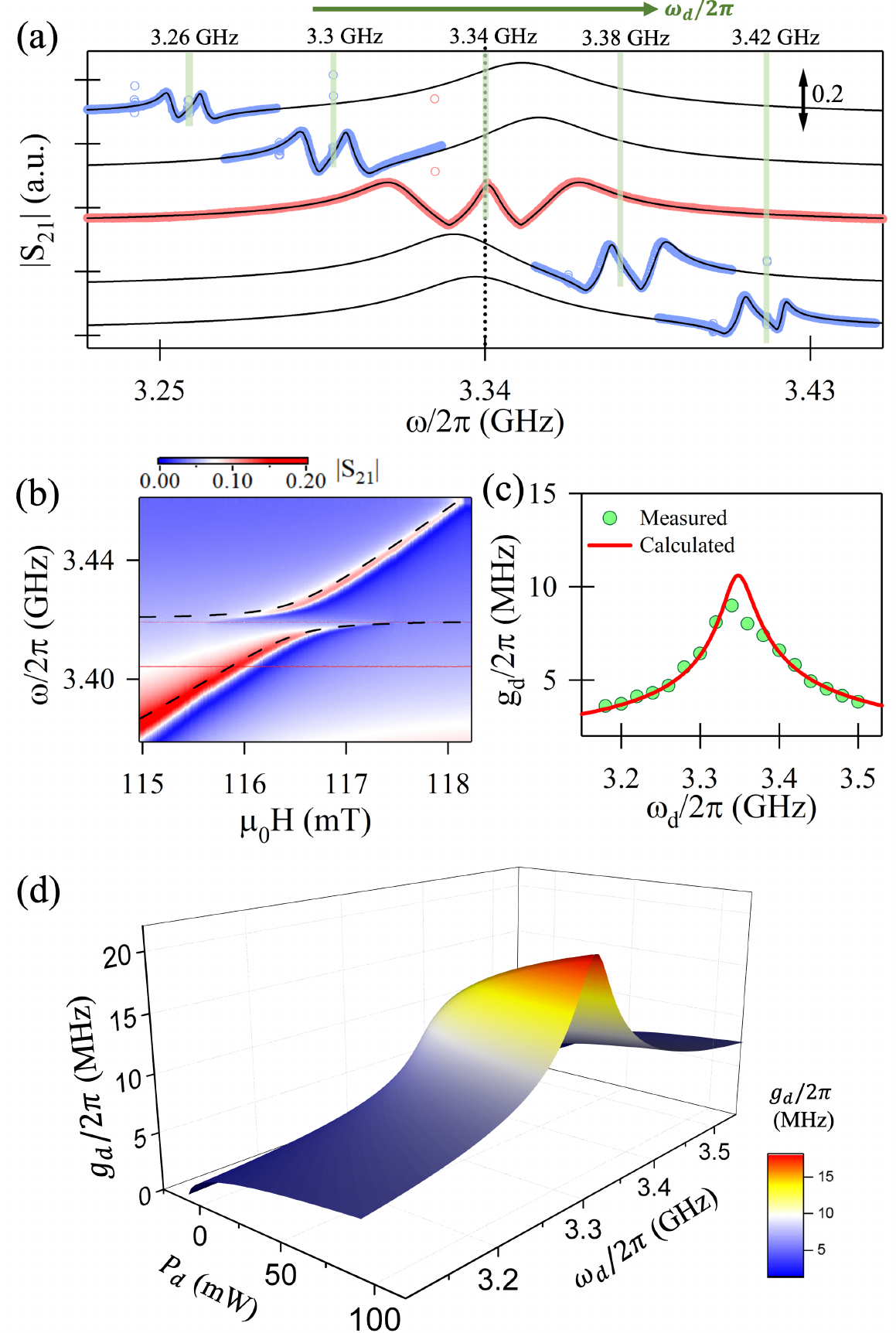,width=8.5cm} \caption{(a) Transmission spectra measured at different pump frequencies. Small coupling gaps appear when the Walker mode matches the pump frequency (i.e., $\omega_w=\omega_d$). Red circles indicate the case of $\omega_w=\omega_d=\omega_c$. (b) Transmission spectra measured at different $H_{ext}$, when setting $\omega_d/2\pi=3.42$ GHz. Black dashed lines are calculated eigenfrequencies from the Hamiltonian. (c) $g_d$ at different pump frequencies extracted from the transmission spectra of panel (a). The red solid line is calculated by using Eq. (\ref{CS}). (d) $g_d$ as a function of $P_d$ and $\omega_d$ and calculated by using Eq. (\ref{CS}). The scattered circles in (a) and the red horizontal line in (b) are spurious signals arising from crosstalk between the microwave generator and the VNA.}\label{fig4}
\end{center}
\end{figure}

Unlike Walker modes, the PIM can be manipulated by the pump. Consequently, the PIM-CMP hybridization inherits this property, and becomes tunable by the pump power and frequency. We fix the Walker mode ($\omega_w$) and the pump frequency ($\omega_d$) to the cavity mode, i.e., $\omega_w=\omega_d=\omega_c$ and then tune the pump power and record the transmission spectrum at each power value. Figure\,\ref{fig3}(a) shows four typical spectra measured at $P_d=$-10, 0, 10 and 19 dBm. The PIM starts to appear at around $P_d=0$ dBm. Three hybridized modes then occur at $\omega_c$ and $\omega_\pm=\omega_c\pm\sqrt{g^2+g_d^2}$. Using Eq. (\ref{Trans}) to fit each transmission spectra, we extract three hybridized modes' frequencies and coupling strengths $g$,$g_d$ and $g_c$, which are plotted in Fig.\,\ref{fig3}(b) and (c). As the pump power increases, the gap between $\omega_\pm$ increases, whereas $g$ remains essentially constant at 24 MHz. Nevertheless, $g_d$ is approximately proportional to the fourth root of the pump power ($g_d\propto P_d^{1/4}$). The physical origin is that the PIM's spin number is equal to the magnons excited by the pump. In another word, the PIM's spin number is altered by the pump power. Such a variation of $g_d$ can be well explained by Eq. (\ref{CS0}), as shown by the red solid line in Fig.\,\ref{fig3}(c). During the calculation, $g_0$ is a constant $5.6\times10^{-4}$, while $A_d$ is approximately set as $\sqrt{P_d}$. $g_c$ is almost zero, which indicates that the direct coupling effect between the PIM and the cavity photon mode is neglectable comparing to two other coupling processes. The physical reason is that the effective spin number of the PIM is much smaller than that of the Walker mode. The measured half coupling gaps at different pump powers, [i.e., $(\omega_+-\omega_-)/2$] are plotted in Fig.\,\ref{fig3} (d) as purple circles. They can be fitted by $\sqrt{g^2+g_d^2}$, as shown by the red solid line in Fig.\,\ref{fig3}(d). At high powers, the experimental data (circles in Fig.\,\ref{fig3}(c) and \ref{fig3}(d)) gradually deviate from the fitting curves (red solid lines) due to the occurrence of nonlinearities that are not considered in this work. \textcolor{blue}{More discussion see Appendix C.} 

Another way to control the PIM-CMP hybridization is changing the pump detuning relative to the cavity mode ($\Delta_c=\omega_c-\omega_d$). The PIM strictly follows the pump frequency, making the anti-crossing of three hybridized modes controllable by the pump frequency. Moreover, the PIM-Walker coupling strength $g_d$ also exhibits a dependency on the pump detuning, because the pump field felt by the YIG sphere that determines the PIM's spin number is regulated by the cavity, as well as its coupling with Walker modes. Figure\,\ref{fig4}(a) shows transmission spectra measured at different pump frequencies [see green lines in Fig.\,\ref{fig4}(a)]. Two resonance peaks appear with a coupling gap, when the CMP mode matches the pump frequency. [blue circles in Fig.\,\ref{fig4}(a)]. Specifically, we fix the pump frequency at $\omega_d/2\pi=3.42$ GHz, and measure the transmission spectra at different $H_{ext}$, as shown in Fig.\,\ref{fig4}(b). The red color in the map indicates the upper branch of CMP modes. An anticrossing appears when the CMP mode approaches the pump at 3.42 GHz, because of the strong PIM-CMP coupling. Black dashed lines are eigenfrequencies calculated from the Hamiltonian (\ref{Hamiltonian}), which are consistent with measurement results. The PIM-CMP coupling strength ($g_d$) is tunable by the pump detuning. It becomes larger, as the pump frequency approaches the cavity mode. $g_d$ fitted from Fig.\,\ref{fig4}(a) are plotted in Fig.\,\ref{fig4}(c). It reaches the maximum when the pump frequency matches the cavity mode, and decreases rapidly as the pump detuning increases. Using Eq. (\ref{CS}), we can calculate $g_d$ at different pump detuning, as shown by the red solid line in Fig.\,\ref{fig4}(c). Furthermore, we also calculate the variation of $g_d$ in a parameter space formed by the pump power $P_d$ and the pump frequency $\omega_d$, as shown in Fig.\,\ref{fig4}(d). Increasing the pump power or decreasing the pump detuning can enhance the effective pump field applied to the YIG sphere. It increases the PIM's spin number $N_p$, and hence enhances the PIM-CMP coupling strength $g_d$. 

\section{Conclusion}

We apply a strong microwave pump on a cavity magnonic system and observe the hybridization between a PIM and CMP modes. This hybrid system inherits intriguing properties from the PIM, allowing it to be controlled by the microwave pump, a feature that is absent in all previous cavity magnonic systems. By using either the pump power or the pump frequency, we can thus tune the PIM-CMP hybridization. In conventional cavity magnonic systems, photon-magnon coupling strength is determined by three parameters, i.e., the spin number of magnets, the mode volume of the cavity photon mode and the overlap coefficient between the cavity photon mode and the magnon mode. This fact causes that the photon-magnon coupling strength is a constant once a cavity magnonic device is fabricated. To adjust the photon-magnon coupling, we must redesign the device or add complicate mechanical structures to change the magnet position in the cavity. Now by utilizing the PIM-CMPs hybridization, we find a handy method that can in situ control the photon-magnon coupling by modulating the pump.Additionally, the parameters, such as the geometrical shape of magnets, the pump duration time and the material properties, are also promising to manipulate the PIM-CMPs hybridization. These methods help to enrich the toolbox for controlling cavity magnonic systems, which should benefit magnon-based signal processing in a broader perspective.

After years of study, CMP generated by the photon-magnon hybridization has become a well-developed technology, and our present work takes a step further. In addition to the formation of CMPs, coherent magnons excited by a pump are strongly coupled with CMPs. From the transmission spectra, we can clear see the anticrossing behaviour arsing from this PIM-CMP hybridization. Normally, anticrossing is an important experimental evidence for the formation of new elementary excitation. We thus conjecture that the PIM-CMP hybridization may generate new elementary excitation that is the combination of two magnons and one photon. This elementary offers a high-dimensional control to realize the photon-magnon entanglement for coherent/quantum information technologies. Our current results just uncover the tip of the iceberg of this elementary excitation and it is expected that more features of non-equilibrium cavity magnonic systems would be discovered.

\acknowledgements

This work has been funded by National Natural Science Foundation of China under Grants Nos.12122413, 11974369, 11991063 and 12204306, STCSM Nos.21JC1406200 and 22JC1403300, the Youth Innovation Promotion Association  No. 2020247 and Strategic priority research No. XDB43010200 of CAS, the National Key R\&D Program of China (No. 2022YFA1404603, 2022YFA1604400), the SITP Independent Foundation, the Shanghai Pujiang Program (No. 22PJ1410700).

\appendix
\renewcommand{\theequation}{\thesection.\arabic{equation}}
\section{Dynamic equations of the PIM-CMP hybridization}
From the Hamiltonian (Eq. (\ref{Hamiltonian})), we can derive the dynamic equations of the PIM-CMP hybridization, which are:
\begin{eqnarray}
    \frac{d\hat{a}}{dt}&=&-i(\tilde{\omega}_w\hat{a}+g_d\hat{b}+g\hat{c}) \nonumber \\
     \frac{d\hat{b}}{dt}&=&-i(\tilde{\omega}_d\hat{b}+g_d\hat{a}+ g_c\hat{c}) \label{A1} \\
     \frac{d\hat{c}}{dt}&=&-i(\tilde{\omega}_c\hat{c}+g\hat{a} + g_c\hat{b}+i\sqrt{\kappa}A_pe^{-i\omega_pt}+i\sqrt{\kappa}A_de^{-i\omega_dt}) \nonumber.
\end{eqnarray}

\vspace{-1mm}
Under the two-tone drive, the oscillation of our system can be viewed as a superposition of a steady oscillation and small fluctuations. The fluctuations are governed by three coupled equations
\begin{eqnarray}
\frac{d\delta u}{dt}&=&-i\tilde{\Delta}_w\delta u-ig_d \delta v-ig \delta\rho\nonumber\\
\frac{d\delta v}{dt}&=&-\xi\delta v-ig_d\delta u -ig_c\delta\rho\\
\frac{d\delta\rho}{dt}&=&-i\tilde{\Delta}_c \delta\rho-ig\delta u -ig_c\delta v+\sqrt{\kappa}A_pe^{-i(\omega_p-\omega_d)t}, \nonumber
\label{A2}
\end{eqnarray}
from which we can obtain the fluctuation of the cavity mode, which is
\begin{equation}
\delta\rho=\frac{i\sqrt{\kappa}A_pe^{-i(\omega_p-\omega_d)t}}{(\omega_p-\tilde{\omega}_c)-\frac{(\omega_p-\tilde{\omega}_d)g^2+(\omega_p-\tilde{\omega}_w)g_c^2+2gg_cg_d}{(\omega_p-\tilde{\omega}_d)(\omega_p-\tilde{\omega}_w)-g_d^2}}.\label{A3}
\end{equation}
Substituting Eq. (\ref{A3}) into the input-output relation of the cavity, i.e., $S_{21}=\sqrt{\kappa}|\delta\rho(t)|/A_p$, we can derive the transmission spectrum of our device, i.e., Eq. (\ref{Trans}). 

\vspace{3mm}
\section{Steady oscillation amplitude of the cavity mode}

According to the experimental results shown in Fig.\,\ref{fig3} (c), the direct coupling effect between the PIM and the cavity photon mode is neglectable, so that we assume $g_c=0$. Then, the steady oscillation amplitudes of three modes derived from Eq. (\ref{A1}) follow
\begin{eqnarray}
\tilde{\Delta}_w A+g_dB+g C&=&0\nonumber\\
-i\xi B +g_d A&=&0\nonumber\\
\tilde{\Delta}_c C+ g A + i\sqrt{\kappa}A_d&=&0
\label{B1}.
\end{eqnarray}
The PIM's spin number is assumed to be the magnon number excited by the pump, i.e., $N_p=\langle\hat{b}^\dag\hat{b}\rangle$, so that the PIM-Walker coupling strength is
\begin{eqnarray}
g_d&=&g_0\sqrt{\langle\hat{b}^\dag\hat{b}\rangle} \nonumber \\
&=&g_0\sqrt{[B^*+\delta v^*(t)][B+\delta v(t)]} \nonumber \\
&\approx&g_0|B|
\label{B2}.
\end{eqnarray}
Substituting Eq. (\ref{B2}) into Eq. (\ref{B1}), we can derive a cubic equation of $B$, i.e., Eq. (\ref{CS}).

\vspace{3mm}
\section{Distinguishing the coupling effect and interference, the coupling regime of $g_d>g$}

\begin{figure} [ht]
\begin{center}
\epsfig{file=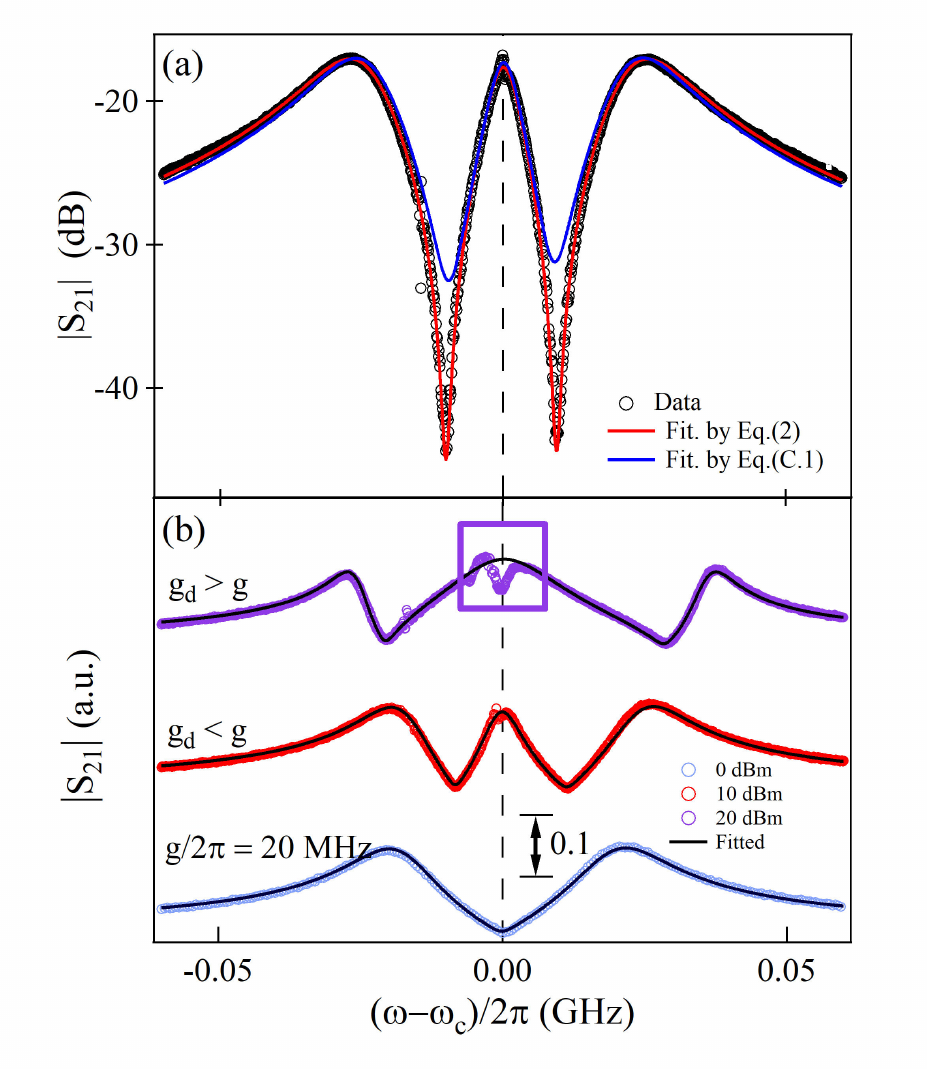,width=8.5cm} {\caption{(a) Fitting the measured transmission spectrum ($|S_{21}|$) by using Eq.(\ref{Trans}) and Eq.(\ref{C1}), respectively. (b) $|S_{21}|$ at the zero detuning ($\omega_w=\omega_c$) in the coupling cases $g_d = 0$, $g_d < g$, $g_d > g$, respectively. Black solid lines are calculated $|S_{21}|$ by using Eq.(\ref{Trans}).}\label{fig5}}
\end{center}
\end{figure}

Because of the presence of the PIM, we can see three resonant peaks in the transmission spectra. Actually, there exist two different mechanisms that can produce this transmission. One is the coupling effect between the PIM and CMPs. The other is the interference effect, in which the PIM is an independent resonance that superposes on the transmission of the two CMP modes. In this case, there is no PIM-CMP coupling effect. Accordingly, we can write the expression of the transmission spectrum caused by the interference, which is
\begin{eqnarray}
&S_{21} &= \label{C1}  \\
&&-\frac{\kappa}{i(\omega_p-\tilde{\omega}_c)+\frac{g^2}{i(\omega_p-\tilde{\omega}_w)}}-\frac{\kappa'}{i(\omega_p-\omega_d)-(\kappa'+\xi)}, 
\nonumber
\end{eqnarray}
where the first term represents the transmission of two CMP modes, and the second term is a single resonance that represents the transmission spectrum of PIM. We use this expression to fit the measured transmission spectrum where the pump power is 10 dBm, but the fitting curve cannot reproduce the measured transmission spectrum well (see Fig.\ref{fig5}(a) blue solid line). The red solid line is the fitting curve by using Eq.(\ref{Trans}). The coefficient of determination for the red line is 0.998, while the coefficient of determination for the blue line is 0.976, $\alpha/2\pi=5.87$ KHz, $\kappa/2\pi=2.33$ MHz and $\kappa'/2\pi=0.42$ MHz. $\alpha/2\pi$ refers to the damping of the magnet and cannot be so close to zero. From the comparison of these curve fitting results, we tend to believe that the PIM-CMP coupling effect does exist, rather than the interference.

In the main text, the PIM-CMP coupling effect locates in the regime of $g_d<g$, but leaves the regime of $g_d\geq g$ unexplored. Increasing the pump power is a feasible method to reach the regime of $g_d\geq g$, because $g$ is independent of the pump power. However, $g_d$ follows a forth root dependence on the pump power, i.e., $g_d\propto P_d^{1/4}$. Therefore, we must use a very high power pump, which will ignite the complicate nonlinearities in magnetic materials. In addition to increasing the pump power, delicately adjusting the YIG position in the cavity can also reach the regime of $g_d\geq g$. Accordingly, we slightly adjust the YIG position to lower the cavity-Walker mode coupling strength to $g/2\pi=20$ MHz from the original 24 MHz. Then we turn on the pump. When the pump power is 10 dBm (i.e., 10 mW), the PIM-Walker mode coupling strength is $g_d/2\pi=9.5$ MHz, still in the regime $g_d<g$. When we increase the pump power to 20 dBm (100 mW), $g_d/2\pi$ increases to 23.7 MHz, which is larger than the cavity-Walker mode coupling ($g$). The measured results of these two coupling cases are respectively shown in Fig.\ref{fig5}(b) with red and purple color, respectively. In the $g_d>g$ regime, besides those three resonant peaks, an additional resonant dip occurs near the pump frequency, as marked by a rectangle in Fig.\ref{fig5}(b). This feature is beyond of our current theoretical model. We conjecture that it may arise from the strong nonlinearity of the Walker mode and the PIM in the $g_d>g$.

\end{document}